\documentclass[10pt,letterpaper]{article}
\usepackage[top=0.85in,left=2.75in,footskip=0.75in]{geometry}

\usepackage{amsmath,amssymb}

\usepackage{changepage}

\usepackage[utf8x]{inputenc}

\usepackage{textcomp,marvosym}

\usepackage{cite}

\usepackage{nameref,hyperref}

\usepackage[right]{lineno}

\usepackage{microtype}
\DisableLigatures[f]{encoding = *, family = * }

\usepackage[table]{xcolor}

\usepackage{array}

\newcolumntype{+}{!{\vrule width 2pt}}

\newlength\savedwidth

\raggedright
\usepackage[aboveskip=1pt,labelfont=bf,labelsep=period,justification=raggedright,singlelinecheck=off]{caption}

\makeatletter
\renewcommand{\@biblabel}[1]{\quad#1.}
\makeatother

\usepackage{lastpage,fancyhdr,graphicx}
\usepackage{epstopdf}
\fancyheadoffset[L]{2.25in}
\fancyfootoffset[L]{2.25in}
\begin{document}
\vspace*{0.2in}

\begin{flushleft}
{\Large
\textbf\newline{The Laplace project: an integrated suite for correlative atom probe tomography and electron microscopy under cryogenic and UHV conditions} 
}
\newline
\\
Leigh T. Stephenson\textsuperscript{1},
Agnieszka Szczepaniak\textsuperscript{1,2},
Isabelle Mouton\textsuperscript{1},
Kristiane A.K. Rusitzka\textsuperscript{1},
Andrew J. Breen\textsuperscript{1},
Uwe Tezins\textsuperscript{1},
Andreas Sturm\textsuperscript{1},
Dirk Vogel\textsuperscript{1},
Yanhong Chang\textsuperscript{1},
Paraskevas Kontis\textsuperscript{1},
Alexander Rosenthal\textsuperscript{3},
Jeffrey D. Shepard\textsuperscript{2},
Urs Maier\textsuperscript{4},
Thomas F. Kelly\textsuperscript{2},
Dierk Raabe\textsuperscript{1},
Baptiste Gault\textsuperscript{1*}
\\
\bigskip
\textbf{1} Max-Planck-Institut für Eisenforschung GmbH, Max-Planck-Stra{\ss}e 1, 40237 Düsseldorf, Germany
\\
\textbf{2} Cameca Instruments Inc., 5470 Nobel Dr, Fitchburg, WI 53711, USA
\\
\textbf{3} Microscopy Improvements e.U., Rudolf von Eichthal str. 66/6,  7000 Eisenstadt, Austria
\\
\textbf{4} Ferrovac GmbH, Thurgauerstrasse 72, 8050 Zürich, Switzerland
\\
\bigskip

* Corresponding author: b.gault@mpie.de

\end{flushleft}
\section*{Abstract}
We present sample transfer instrumentation and integrated protocols for the preparation and correlative characterization of environmentally-sensitive materials by both atom probe tomography and electron microscopy. Ultra-high vacuum cryogenic suitcases allow specimen transfer between preparation, processing and several imaging platforms without exposure to atmospheric contamination. For expedient transfers, we installed a fast-docking station equipped with a cryogenic pump upon three systems; two atom probes, a scanning electron microscope / Xe-plasma focused ion beam and a N\textsubscript{2}-atmosphere glovebox. We also installed a plasma FIB with a solid-state cooling stage to reduce beam damage and contamination, through reducing chemical activity and with the cryogenic components as passive cryogenic traps. We demonstrate the efficacy of the new laboratory protocols by the successful preparation and transfer of two highly contamination- and temperature-sensitive samples - water and ice. Analysing pure magnesium atom probe data, we show that surface oxidation can be effectively suppressed using an entirely cryogenic protocol (during specimen preparation and during transfer). Starting with the cryogenically-cooled plasma FIB, we also prepared and transferred frozen ice samples while avoiding significant melting or sublimation, suggesting that we may be able to measure the nanostructure of other normally-liquid or soft materials. Isolated cryogenic protocols within the N\textsubscript{2} glove box demonstrate the absence of ice condensation suggesting that environmental control can commence from fabrication until atom probe analysis.

\section*{Introduction}
Atom probe tomography (APT) is now an essential analytical tool for advancing modern material science applied to engineering materials \cite{Devaraj2018}, the life sciences \cite{Adineh2016} and geology \cite{Valley2015}. Experimentation remains challenging, chiefly in fabricating undamaged APT samples and, once made, transferring those samples between microscopes without environmental contamination or modification. Addressing these difficulties could facilitate many unique studies in materials, surface sciences and, possibly, the life sciences. 

Amongst the issues regularly faced is the analysis of highly reactive metals: the bare surface of a fresh specimen tends to rapidly react with the gaseous environment and hence what gets analysed is not the actual material of interest any longer. Beyond difficulties inherent to preparing specimens for surface analyses, environmental degradation of a specimen’s surface and sub-surface region has hindered exploiting of the full promise of APT for the analysis of catalysts despite some valiant efforts in this space \cite{Bagot2006,Li2018} using \textit{in-situ} techniques \cite{Barroo2014}. Recent innovations for sample transfer methods permit the characterization of environmentally sensitive materials \cite{Perea2017}. Deuterium distribution in a ferritic steel \cite{Chen2017,Takahashi2018} and a palladium alloy \cite{Haley2017} were isolated with the development of cryogenic transfer protocols. A novel experiment preparing liquid has been assisted by the use of a modified Leica cryogenic suitcase at ETH, Zurich \cite{Gerstl2015,Dumitraschkewitz2016}. Although in development for several years, these approaches are far from routine and existing instrumentation usually maintains the specimens under high-vacuum conditions (approx. $10^{-6}$ mbar), which may be above the partial pressure necessary to activate surface reactions (e.g. oxidation). 

The comprehensive modular protocol developed at the Max-Planck-Institut für Eisenforschung (MPIE) and presented here caters to multiple research streams requiring these capabilities to be carried out simultaneously. A modular approach is employed where samples are kept isolated from the environment during transport through the use of a cryogenic ultrahigh vacuum (UHV) carry transfer suitcase (CTS), matching the  ultrahigh vacuum (below $10^{-9}$ mbar) and the low-temperature capabilities of experimental platforms. This UHVCTS enables the movement of specimens under ultrahigh vacuum conditions and with cryogenic temperatures, transferring between the various preparation and analytical platforms that are equipped with cryogenically-cooled stages. 

In the cryo-UHVCTS, the low temperature is maintained through liquid nitrogen (LN2). It was recently shown that maintaining a specimen ``cold chain" transfer at LN2 temperature is sufficient to immobilise a detectable amount of deuterium within the structure of a steel specimen upon electrochemical charging \cite{Chen2017}. This aspect is critical to enable hydrogen mapping, which is one of the major opportunities for APT \cite{ Cairney2017, Takahashi2018}. Hydrogen is the smallest of all atoms, is very mobile, and at room temperature the desorption rate from steels can be significant. For the sample preparation and nanocharacterisation of hydrogen-charged alloys, two UV laser-assisted atom probes, including one with a reflectron, and a xenon-plasma focused-ion beam (PFIB) microscope were equipped to dock with the UHV suitcase. Cryogenic cooling on the PFIB stage additionally allows for sample preparation with less defect agglomeration introduced by ion knock-on damage, radiolysis and heating \cite{BASSIM2012} all of which can be suppressed by the use of lower beam energies, lower currents and sample cooling \cite{LEE2012,Kim2011,Dolph2014}. We hypothesize that sample cooling can also reduce the contamination of the surface and sub-surface volumes, whether by chemisorbed gases (typically H\textsubscript{2} and H\textsubscript{2}O) diffusing into the bulk or via oxidation (absorbed O\textsubscript{2}). Sample cooling also prevents the sublimation of some materials which on the nanoscale can completely destroy the targeted structure. For controlled processing and electrolytic hydrogen/deuterium-charging, as well as other experiments sensitive to the exposure to atmospheric moisture or oxygen, a dedicated N\textsubscript{2} glovebox was also interfaced. We present here the first results obtained with this setup.

Solving the materials design problems presented by metallic materials in hydrogenating environments hinge upon understanding hydrogen's role in the various phenomena which may exacerbate or even directly cause catastrophic failure. Characterisation of nanostructural features, like grain boundaries, stacking faults, dislocations, vacancies, voids, small precipitates and hydrides, is necessary to obtain insight into how hydrogen diffuses through and can be sequestered by these features. Atom probe microscopy can serve as a high-throughput technique for this purpose, as it gives both three-dimensional structural and chemical information with near-atomic resolution. Unfortunately, experiments concerning hydrogen in these materials face challenges that other common approaches employing atom probe do not face \cite{Haley2017,Chang2018}.

Over the past few decades, atom probe tomography has evolved to provide compositionally-resolved three-dimensional structure on sub-nanometre scales within comparatively large volumes (typically 100 x 100 x 750 nm$^3$). Practical atom probe requires the refinement of many steps and proponents of atom probe still grapple with basic concerns regarding data validity and interpretation. The question ``to what extent does an atom probe measurement reflect reality and how can this be improved?" often requires input from additional microscopy techniques. In answer to this, electron microscopy techniques cannot be used by themselves to retrieve all desired information, as their associated compositional analysis is limited and often confined to two dimensional images. Significant efforts have been devoted to establishing correlative microscopy protocols that lead to a more complete understanding of the microstructure of an atom probe specimen \cite{Herbig2018}. Transmission electron microscopy has been used to image atom probe specimens before \cite{Mouton2017} or after acquisition \cite{Petersen2009,Haley2011}, and some efforts characterising both before and after \cite{Kirchhofer2015,Stokes2017}. Electron microscopy in conjunction with atom probe tomography has, in several cases \cite{Arslan2008,Baik2013,Devaraj2014,Herbig2014,Stoffers2017,KwiatkowskidaSilva2017,Kuzmina2015,Makineni2018}, even been applied to atomic-scale resolution TEM imaging of complex microstructures with full chemical resolution.

These solutions are potentially better than palliative strategies that correct for obvious reconstruction flaws. The apparatus described herein demonstrates that UHV transport of specimens (with or without cryogenic preservation), can be an effective component of correlative studies by preventing environmental alteration of the atom probe specimen. The possibility to perform back-and-forth transfers though the cryo-UHVCTS enables imaging of the specimen at various stages of the analysis, which enables recording of the emitter shape. 

Another expected outcome of the project, and a future direction of this research, is the exploitation of this information to inform improved data reconstruction protocols and enable mapping of the very positions of atoms and the chemical compositions at the atomic level in three-dimensions \cite{Kelly2013}, hence the allusion to the ‘Laplace demon’. Simon Pierre Laplace indeed postulated that knowing the position and nature of all particles in the universe at a given point in time, one could predict the future and know all of the past. Modern quantum mechanics has supplanted such perspectives, but ambitions of understanding material evolution still require accurate measurements. Here, we describe the necessary instrumentation to further these ambitions and showcase some of the preliminary results obtained that exploit these novel capabilities.

\begin{figure}[!ht]
\includegraphics[width=\textwidth]{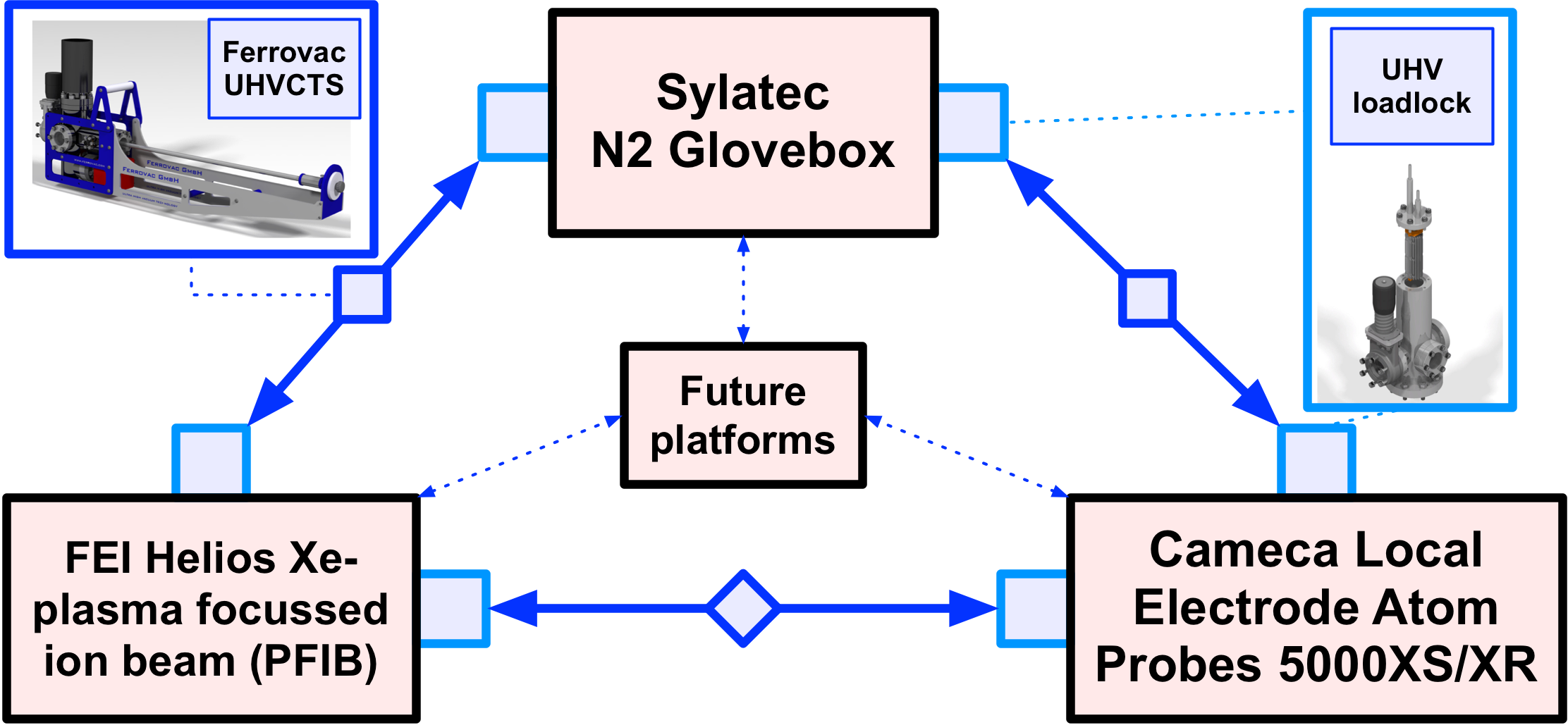}
\caption{{\bf Experimental overview of the Laplace Project.\\}
The ultrahigh vacuum carry transfer suitcase (UHVCTS - inset left) can be securely fastened to every experimental platform via an ultrahigh vacuum loadlock (inset right) which is pumped as necessary to maintain conditions in either microscope or glovebox. It is planned to mount the suitcase upon future instruments. 
}
\label{Fig 1}
\end{figure}

We describe in detail our new suite of instruments and the modifications necessary to enable cryogenic UHV sample transfers. We itemize each individual component and its utility, acknowledging that different designs and alternate off-the-shelf components may provide similar capabilities. Fig~\ref{Fig 1} depicts the interconnectivity between our experimental platforms that the new UHV sample transfer protocols provide. Multiple exchanges can be performed upon the same specimen for investigations requiring additional fabrication, chemical treatment or correlative electron microscopy.

\subsection*{Ultra-high vacuum carry transfer suitcase (UHVCTS)}

The key enabling technology is the two UHVCTS units (Ferrovac VSN40S). Their internal assembly is cooled by liquid N\textsubscript{2} (LN2), and each is modified to accept a modified CAMECA atom probe puck (detailed below). Each UHVCTS has a 500-mm wobblestick that ends with a PEEK-insulated puck manipulator. The two cryo-UHVCTS only differ in their small ion pumps with non-evaporable getter (NEG) cartridges, rated to pump 100 L/s and 200/s respectively (NexTorr D-100-5 vs D-200-5). With proper use, each can easily attain $10^{-10}$ mbar. The UHVCTS are mounted on the experimental platforms via specially designed loadlocks (Ferrovac VSCT40 fast pump-down docks), pumped by a 80 l/s turbopump (Pfeiffer HiPace 80). The contained cryogenic pump is used either when minimizing loading time is essential or for improving the vacuum quality via cold trap action. Figure~\ref{Fig 1} depicts both the UHVCTS and the UHV booster loadlock. 

\subsection*{Atom probe and loadlock}
Two state-of-the-art commercial APT microscopes are available; namely a straight-flight-path instrument, the Cameca LEAP 5000XS, and a reflectron-fitted instrument, the LEAP 5000XR. A docking station was added to both LEAPs. 

APT specimens are mounted upon easily manipulated specimen pucks. For the Laplace Project, we used specially designed pucks that are thermally insulated from any direct contact with vacuum transfer rods or wobble-sticks through a 2mm-thick layer of polyether ether ketone (PEEK) that replaces a part usually made of metal. PEEK is a UHV-compatible polymer that has good strength, low thermal diffusivity, and can be baked. Coupon clip-holders are screwed into the puck from the front of the puck, thereby allowing a slim profile which is advantageous for multiple transfers through a maze of vacuum chambers.

\begin{figure}[!ht]
\includegraphics[width=\textwidth]{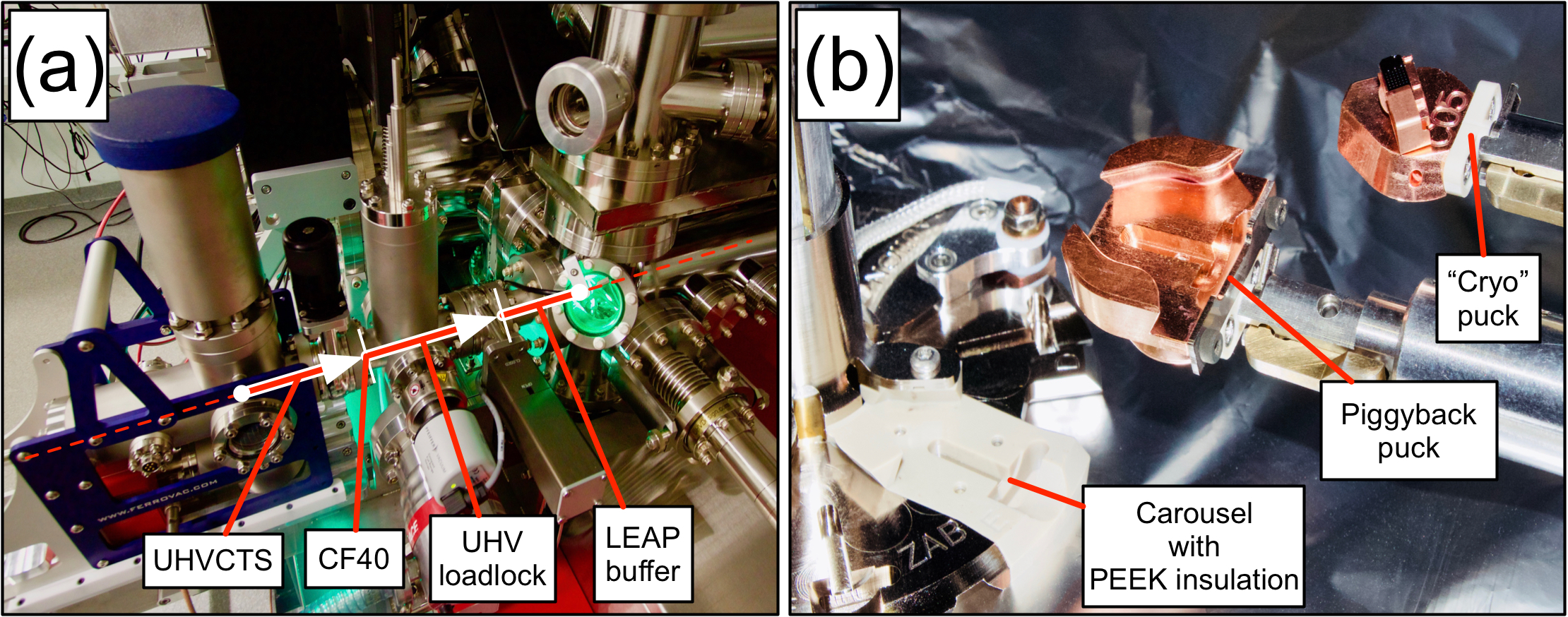}
\caption{{\bf Atom probe modifications.\\}
(a) A side-view of the cryogenic ultrahigh vacuum transfer into the buffer chamber of the local electrode atom probe (LEAP 5000XR). The connection is made at the labelled CF40 flange. (b) An expanded view of the interlocking pieces used to transfer cryogenically-cooled samples from the suitcase to the atom probe's analysis stage with no increase in temperature. (\textit{CF40}: ConFlat\textsuperscript{TM} 40 mm flange; \textit{UHV}: ultrahigh vacuum; \textit{CTS}: carry transfer suitcase; \textit{PEEK}: polyether ether ketone.)}
\label{Fig 2}
\end{figure}

One of the positions in the carousel used to store specimen pucks in the airlock and intermediate buffer chamber of the LEAPs was also modified and is thermally insulated through the use of PEEK. An additional component for rapidly loading a sample puck onto the cryogenic analysis stage was developed. A piggyback puck with a large thermal mass is pre-cooled, typically to a temperature below 50K, by placing it on the cryo-stage of the analysis chamber. The sample puck is passively cooled by this piggyback puck placed into this insulated position on the carousel. This makes it possible to  maintain the transferred specimen puck at low temperature between the transfer from the cooled-down suitcase to the analysis position in the microscope. This is important for temperature-sensitive samples or samples that could be chemically-unstable even at liquid N\textsubscript{2} temperatures. Their assembly are shown in Figure~\ref{Fig 2}(a).

\subsection*{Dual-beam Focussed ion beam modifications}
\begin{figure}[!ht]
\includegraphics[width=\textwidth]{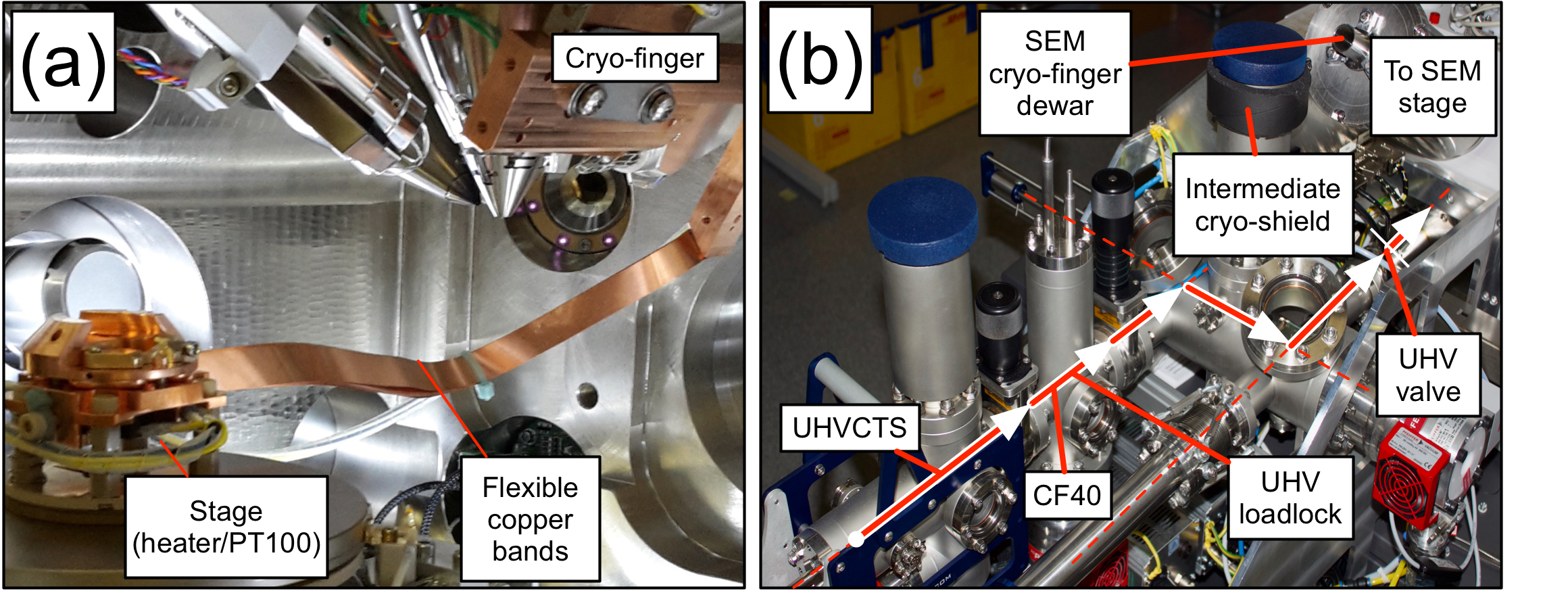}
\caption{{\bf Xenon plasma focussed ion beam microscope modifications.\\}
(a) The configuration of the solid-state cooling bands in the scanning electron microscope (SEM) chamber with PT100 resistance thermometers. (b) An aerial view of the SEM side-chamber allowing for the cryogenic sample transfer between the two different vacuum regimes of the suitcase and the SEM chamber. The connection is made at the labelled CF40 flange (ConFlat\textsuperscript{TM} 40 mm). \textit{UHV}: ultrahigh vacuum; \textit{CTS}: carry transfer suitcase.}
\label{Fig 3}
\end{figure}

Figure~\ref{Fig 3} depicts a schematic for the specimen exchange on the xenon-plasma focussed ion beam (PFIB) microscope. The suitcase is attached via a VSCT40 fast pump-down loadlock to an intermediate buffer chamber with optional cryogenic cooling (via LN2 dewar) on the puck transfer shuttle (-130 $^\circ$C) and a surrounding cryo-shield. The shuttle is used to transport up to two sample pucks between suitcase and SEM chamber insertion positions. The microscope chamber only achieves a high vacuum ($5\times 10^{-7}$ mbar) and this intermediate transfer protocol serves to maintain the suitcase's ultra-high vacuum (typically $\approx 5\times 10^{-10}$ mbar). The buffer has a quick loadlock gate for the insertion of pucks upon the shuttle if required.

The plasma FIB has a bespoke SEM stage which accepts atom probe pucks and is thermally isolated from the stage mount by polyether ether ketone (PEEK) plastic. We can cool the stage with flexible copper bands attached to a nickel/gold-coated copper cold-finger cooled to -184 $ ^\circ$C degrees by an external LN2 dewar. We measure temperature on the stage and at two points on the cold finger with standard PT100 resistive temperature sensors with the minimum stage temperature being approximately -140 $^{\circ}$C after three hours. Next to each sensor, a MOSFET chip can be used to provide rapid heating of the cold finger. Controlled heating can be implemented for sublimation or recrystallization studies.

\subsection*{Glovebox}
The two UHVCTS are mountable on a modified Syletec glovebox. For this instrument, the UHV loadlock has been augmented with a Agilent Starlab ion pump (isolated from high pressures by a UHV valve) and the venting performed with ultra-clean N\textsubscript{2}. For specimen transfer, the UHVCTS is opened directly to a clean N\textsubscript{2} glovebox atmosphere  with 0.5 ppm O\textsubscript{2} and a dew-point $< -105$ $^{\circ}$C (below the measurement limit). The UHVCTS vacuum adequately recovers with the support of the pumping system on the fast-docking station as well as the additional ion pump. The UHVCTS ion pump can then maintain $\approx 10^{-8}$ mbar but the ion pump's non-evaporable getter must be reactivated to restore optimal performance.

\section*{Results}
\subsection*{Metallurgical examples}
\begin{figure}[!ht]
\includegraphics[width=\textwidth]{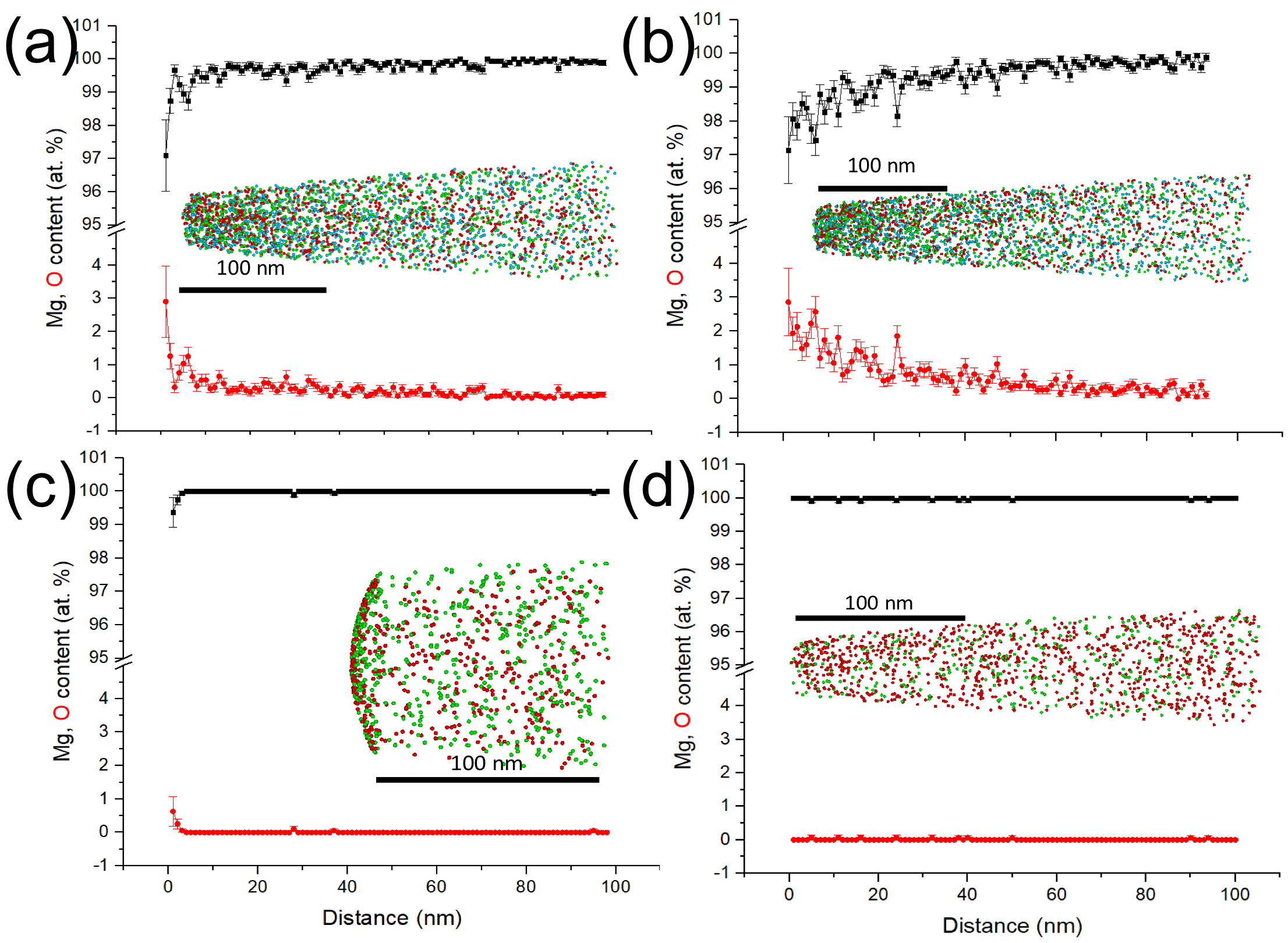}
\caption{{\bf Preparation and ultrahigh vacuum transfer of magnesium.\\} Atom probe reconstructions of pure magnesium samples prepared and transferred in the following protocols. (a) Room temperature fabrication, stored for 1 week in buffer, removed to atmosphere for 3 minutes before the atom probe. (b) Room temperature fabrication, stored for 2 weeks in buffer before atom probe. (c) Milled using the cryogenically-cooled stage, stored for 1 week in buffer before atom probe. (d) Milled using the cryogenically-cooled stage, transferred with the cryogenic ultrahigh vacuum carry transfer suitcase, and then atom probe immediately).
}
\label{Fig 4}
\end{figure}

Pure magnesium samples were prepared in the Helios PFIB, using a 30kV Xe-ion accelerating voltage for rough milling and an 8kV accelerating voltage for fine milling. Each sample was transferred via the UHVCTS to the LEAP 5000XR using  with slightly different protocols. Figure~\ref{Fig 4} shows four reconstructions from atom probe measurements with corresponding compositional profiles running through the specimen. Up to 4 at. \% oxygen was detected in the surface layers when cryogenic protocols was not obeyed at all, i.e. room temperature milling, exposure to atmosphere and prolonged storage in the LEAP5000XR buffer. In the case where the cold stage was used, a cryogenic UHV transfer was made. The sample was run immediately and negligible levels of surface oxygen were detected.

Processing or preparing specimens within the controlled environment inside the glovebox is also expected to allow for an additional level of control over the formation of spurious surface species. This is particularly true for the development of frost or ice that develops on the surface of specimens after electrochemical charging with e.g. deuterium used as a proxy for hydrogen in some recently reported studies\cite{Chen2017, Haley2017}. Upon charging, specimens are quenched and kept in LN2 to prevent further migration of the deuterium, and when this is done in air, the moisture condenses onto the specimen's surface and prevents its direct analysis. Performing these tasks in the glovebox is expected to alleviate or limit the influence of such issues. 

\subsection*{Glovebox isolation and aqueous solutions}
\begin{figure}[!ht]
\includegraphics[width=\textwidth]{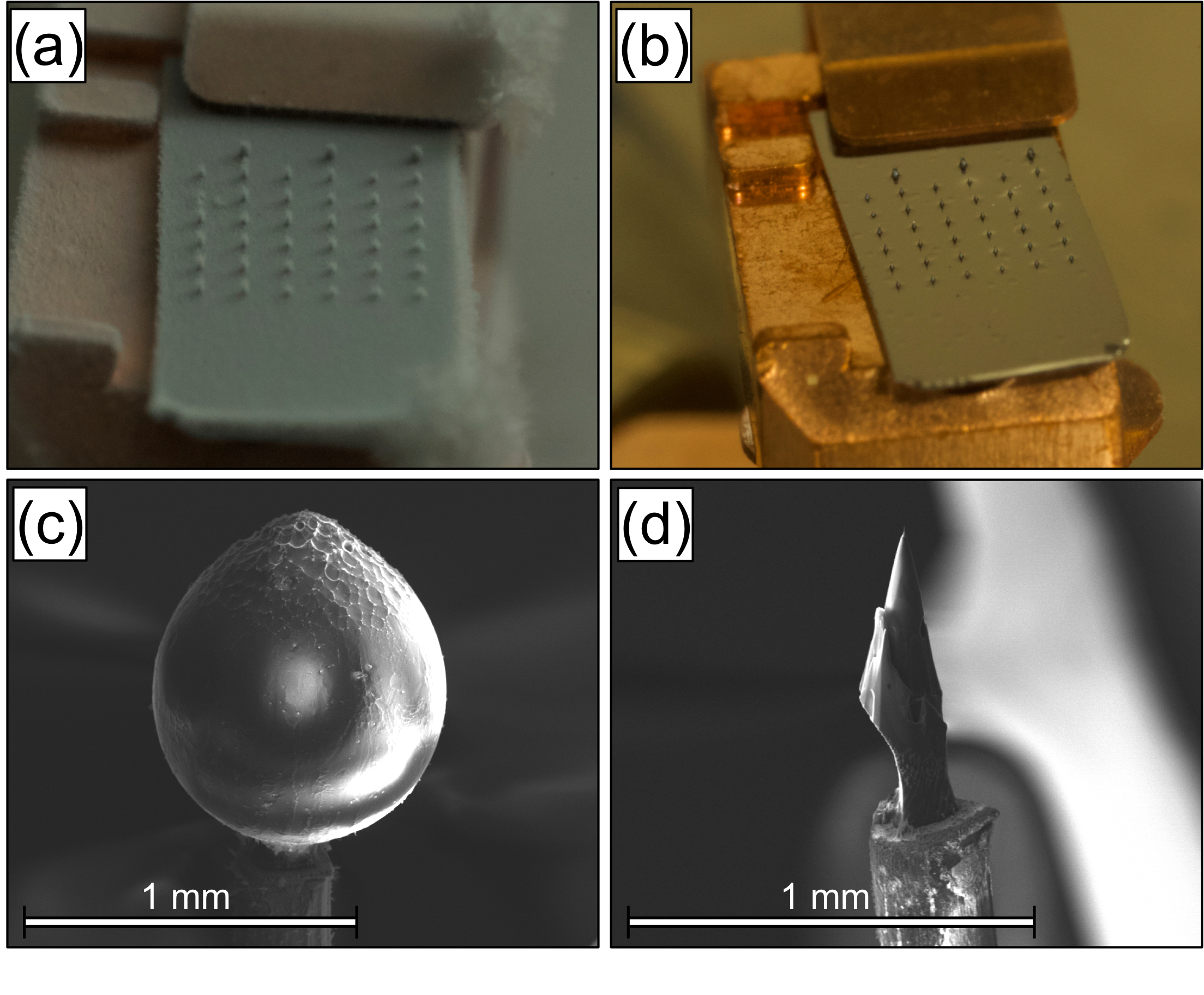}
\caption{{\bf Condensation and ice samples.\\}
(a) A 36-microtip array photographed approximately 90 seconds after removal from a liquid nitrogen bath ($\approx-196~^\circ$C) and evidencing significant ice condensation. (b) The same 36-microtip array, 90 seconds after removal from the liquid nitrogen bath, this time showing no ice condensation. (c) A $\approx$1-mm sphere of pure water ice upon a 0.3-mm stainless steel wire. (d) From the sphere, a 50-nm atom probe needle was fashioned in the plasma focussed ion beam microscope with the cold-stage and transferred to the local electrode atom probe (LEAP 5000XS) with the cryogenic ultrahigh vacuum carry transfer suitcase.}
\label{Fig 5}
\end{figure}

A good ``cold-chain" protocol as presented in \cite{Chen2017} is necessary, not only in introducing a specimen to an atom probe chamber without changing the specimen, but also to minimise contamination of condensed atmospheric gases. Figure \ref{Fig 5}(a) and (b) show the accumulation of moisture condensation at approximately 90 seconds after removal from an LN2 bath upon silicon microtip coupon specimens and an electropolished metallic specimen respectively. Using the glovebox, we eliminated the formation of ice over the same time period after removal from an LN2 bath (Figure \ref{Fig 5}).

Microcopic analyses of soft materials often implies cooling them to cryogenic temperatures, and often vitrifying them as required, so as to minimise modification of the structure during the preparation of the specimens. Techniques related to atom probe have been used to investigate thin layers of vitreous specimens as reported by \cite{Stintz1991,Stintz1992}. However, their use has been rather limited. A protocol to prepare specimens with dimensions suitable for atom probe was tried. First, a drop of water (boiled to remove dissolved gases and doped with $\approx$55 p.p.m. NaCl to increase conductivity) was micro-pipetted upon the flat-top of a 0.3-mm stainless steel wire mounted into a typical puck, pre-cooled with liquid nitrogen and loaded through the quick load-lock chamber attached to the PFIB. The puck was then transferred onto the dedicated pre-cooled stage inside the main chamber of the PFIB. The frozen drop was first cut into a cuboidal shape using the Xe-plasma beam (1.6 nA @ 30kV), and subsequently sharpened into a needle by using a two stage annular milling at 30kV. These successive steps resulted in producing an APT needle as displayed in \ref{Fig 5}(d). Laser-pulsed APT analysis of such a thick layer of ice could not be conducted, likely due to ice's extremely low electrical conductivity. The entire range of high voltage up to 15 kV, and laser pulse energy up to 1 nJ was spanned, but no ions were detected.

\section*{Discussion}
Specimen preparation has often been regarded as the most challenging aspect of performing meaningful APT experiments. An extension of this is to consider transfer and even the atom probe as being part of this preparation process. Exercising higher levels of control over the specimen fabrication and transport will affect both experimental yield and data quality. Moreover, field evaporation is a gradually eroding process which makes atom probe experiments destructive; acquiring microscopic snapshots of APT specimens during this process is crucial to better understand field evaporation and to formulate more accurate three dimensional reconstructions. 

Such observations can only be performed when specimens are held in and transferred through controlled environments, preventing any surface transformation that, at the approximately $<100$-nm scale of an atom probe specimen, can be a substantial part of the analysed specimen volume. The intricacies of specimen preparation, the protocol of specimen transfer and even the routine of specimen storage must be examined. For example, preparation of samples susceptible to O\textsubscript{2}, moisture, or residual chamber gases (even down to $10^{-10}$ mbar) is problematic, so it can be imagined that a new generation of atom probe experiments will focus on such materials sensitive to parameters like temperature and atmospheric exposure. Our results demonstrated that not only should the atmosphere around the specimen be controlled during the transfer of the specimens, but also during the preparation itself.

By use of a cryogenic cold stage in our plasma FIB, and by maintaining cryogenic temperatures under UHV transfer, we demonstrated in the analysis of a magnesium alloy (Figure \ref{Fig 4}) that we can successfully suppress surface oxidation and thus sub-surface modification. Comparing two Mg samples fabricated at room temperature (Figure \ref{Fig 4}(a-b)), where both samples were transferred by UHVCTS and one was briefly removed into atmosphere, we demonstrated that storing the sample in the vacuum chamber was no guarantee that oxidation would not occur (for this particular material at least). Even considering the results for the cryogenically-prepared samples (Figures \ref{Fig 4}(c-d)), we could suppose that longer storage times can still influence surface oxidation. This is not surprising but we suggest that best practice is immediate experimentation of cryo-UHVCTS samples. The complete chain of specimen preparation and handling must thus be controlled.

How does using the cryogenically-cooled stage in the plasma FIB result in less oxygen being detected? As the literature suggests, cooling the specimen can reduce beam-induced damage. Sample cooling may also prevent the radiolysis of adsorbed chamber gases, the radicals of which can then react and transform the specimen's surface. A solution could also be offered through unintentional cryo-pumping. In the setup in Figure \ref{Fig 3}, the cold finger and the copper bands present a large surface area and so could provide a large cold surface for H\textsubscript{2}O, CO\textsubscript{2} and other oxygen-containing hydrocarbons to condense upon. The cold finger reaches -184 $^\circ$C, and though molecular oxygen condenses at -183 $^\circ$C at standard pressure and temperature, it would not condensed at the vapour pressures present in the plasma FIB. This could suggest that better vacuum management could improve specimen preparation.

We shall soon make minor modifications to the plasma FIB stage, addressing the efficiency of the stage cooling but also the frequent failure of the MOSFET heater chips, likely replacing these in the future with small resistive heaters. This will increase the speed of specimen preparation but also allow for specific temperatures to be reached upon the stage. One useful application  would be to allow the platinum precursor, introduced by the plasma FIB's gas injection system, to be used at non-cryogenic temperatures $<0~^\circ$C. A low cryogenic temperatures, this precursor condenses upon the cooled specimen without site specificity. Another use for variable temperature will be for sublimation control of an aqueous substrate.

We have proven that the N\textsubscript{2} glovebox provides a workspace for experiments requiring isolation from the atmosphere. Such experiments may include electrolytic hydrogen charging of materials where the sample can then be subsequently plunged to liquid nitrogen temperatures. Figure \ref{Fig 5}(b) demonstrates that water vapour is scrubbed from the glovebox to negligible levels. With the glovebox, we will be able to safely handle oxygen-sensitive materials like hydrides for energy applications or reactive alkali metals. Thereafter, introducing such samples to the closed UHVCT protocols (with possible cryogenic capabilities) will allow for atom probe investigations which have previously not been considered realistic.

\section*{Conclusions}
We have demonstrated the basic functionality and effectiveness of our ultrahigh vacuum carry transfer suitcases and associated protocols. The level of control of the environment of the specimen pre- and post-preparation, up until the actual atom probe analysis, is unprecedented. We demonstrate that application of our protocols can alleviate the formation of an oxide layer formed on the surface of a reactive metal such as Mg, and also that we can reduce the depth of penetration of spurious O that can extend over tens of nanometers below the surface. This should greatly improve O-impurity measurements in such systems. We expect that the use of such cryo-protocols will become more widely spread in the coming years.  

\section*{Acknowledgements}
The Bundesministerium für Bildung und Forschung is acknowledged for the funding of the UGSLIT project and the Max-Planck Gesellschaft for the funding of the Laplace Project. AKR is grateful for funding from the Volkswagen Stifftung through the Experiment! scheme. YC is grateful for financial support from the Chinese Scholarship Council. Sigrun Köster from Ferrovac is acknowledged for her support to the project and fruitful discussions. We are also grateful to Dr. Abhishek Tripathi for providing the magnesium samples and Laila Moreno Ostertag for concocting a dilute saline solution that we used for the preparation of our ice atom probe specimen.

\nolinenumbers

\end{document}